\documentclass[aps,prl,twocolumn,amssymb]{revtex4}

\usepackage{graphicx}

\usepackage{CJK}
\usepackage{ulem}
\usepackage[T1]{fontenc}
\usepackage{calligra}

\usepackage{graphicx}
\usepackage{dcolumn}
\usepackage{bm}

\usepackage{times}
\usepackage{verbatim}
\usepackage{graphicx}
\usepackage{graphics,epsfig}

\usepackage{theorem}
\usepackage{makeidx}
\usepackage{amsmath}
\usepackage{epic}
\usepackage{amscd}

\usepackage{bbm}
\usepackage{xy}
\usepackage{amssymb}
\usepackage{epsfig}

\begin{document}

\pagestyle{empty}
{\bf \noindent Comment on ``Energy Transfer and Dual Cascade in Kinetic Magnetized Plasma Turbulence''}
\vspace{0.2cm}

Some important inappropriate physical statements and some mathematical mistakes in the Letter \cite{GabeTomoPRL11} by Plunk and Tatsuno (LPT) (who made the contribution in trying to establish the transfer constraints of gyrokinetics) are pointed out. The Fj{\o}rtoft \cite{Fjortoft} constraints that LPT presented (but not correctly), do not predict the transfer directions which however may be assisted by the corresponding absolute equilibria calculated in this Comment, following Kraichnan \cite{KraichnanDualCascade}.

With $\hat{}$ and $\breve{}$ indicating respectively Fourier and Hankel transform, LPT starts from their Ref. [7]
\begin{eqnarray}\label{eq:gk}
\frac{\partial \breve{\hat{g}}(\textbf{k},b)}{\partial t}&=&\int_0^{\infty} vdv J_0(bv) \textbf{z}\times \sum_{\textbf{p}+\textbf{q}=\textbf{k}} \textbf{p}J_0(pv)\hat{\varphi}(\textbf{p})\cdot \nonumber \\
&& \textbf{q} \int_0^{\infty} wdw J_0(vw)\breve{\hat{g}}(\textbf{q},w)
\end{eqnarray}
which expresses that, with the quasineutrality condition $\hat{\varphi}(\textbf{p})=\beta(\textbf{p})\breve{\hat{g}}(\textbf{p},p)$, the dynamics must involve at least one "diagonal" mode $\breve{\hat{g}}(\textbf{p},p)$, otherwise $\frac{\partial \breve{\hat{g}}(\textbf{k},b)}{\partial t}=0$; and, when only one of the modes is diagonal, this diagonal component is frozen, only mediating the $\textbf{k}$ and $\textbf{q}$ modes. So, LPT's claim, in the last paragraph of the second page, about the transitions involving only one or no diagonal components is invalid.

Suppose velocity is bounded by $V$, we then have $\hat{g}(\textbf{k},v)=\sum_{z} 2V^{-2}[J_1(z)]^{-2}\breve{\hat{g}}(\textbf{k},z) J_0(z v/V)$, with $z$ being the zeros of $J_0$. LPT takes the upper bound $V$ to be $\textbf{k}$ dependent so that for each $k$ there is some zero $z_k$ of $J_0$ such that $k=\frac{z_k}{V(\textbf{k})}$. Such a courageous step will introduce subtleties, which we would not elaborate here but just expose the very obvious and direct point which annoys LPT and was treated wrongly: When the upper bound of the integral over $v$ depends on $\textbf{k}$, one can not change the order of the integral over $v$ and the sum over $\textbf{k}$, and that $V(\textbf{k})$ could not be pulled out and normalized as done in LPT's Eq. (5). Of course, if one could (most probably not always) find a uniform upper bound for all, even finite number of, $\textbf{k}$, things would work; but, this is not considered by the authors. Another subtlety is that the $\textbf{k}$ dependence of cutoff could also affect the relation between the spectra of $E$ and $W$ as we will point out after we present the corresponding ($v$-bounded) system's absolute equilibria below.

In the scale, $k$-$b$, space, the invariants used in LPT are $E=\sum_{\textbf{k}}\beta(\textbf{k})\int_0^{\infty} \delta(k-b)|\breve{\hat{g}}(\textbf{k},b)|^2db$ and $W=\sum_{\textbf{k}} \int_0^{\infty} |\breve{\hat{g}}(\textbf{k},b)|^2bdb$. The rugged Fourier-Bessel Galerkin truncated invariants corresponding to those applied by LPT are
$\tilde{E}=\tilde{\sum}_{\textbf{k}}\tilde{\sum}_z \frac{\pi}{2} \beta(\textbf{k}) \delta_{z,z_k}|\breve{\hat{g}}(\textbf{k},z)|^2$ (with $z_k=kV(\textbf{k})$, and, $\delta_{z,z_k}$ acquires 1 for $z=z_k$ and $0$ otherwise,)
$\tilde{W}=\tilde{\sum}_{\textbf{k}}\tilde{\sum}_{z} 2V^{-2}(\textbf{k})J_1^{-2}(z)|\breve{\hat{g}}(\textbf{k},z)|^2$, where $\tilde{\bullet}$ means operating only on a subset, that is the Galerkin truncation as in Ref. \cite{ZhuHammettPoP2010}. Comparing the densities (here are those kernals behind $\tilde{\sum}_{\textbf{k}}\tilde{\sum}_{z}$,) we see that LPT's Eqs. (1, 5 and 6) are not appropriate.

Note that to study the transfers of the global invariants in velocity scale space, now one can not take all the local (in $v$) invariants $G(v)=\tilde{\sum}_{\textbf{k}}|\hat{g}(\textbf{k},v)|^2$, discrete or not \cite{ZhuHammettPoP2010,ZhuDualCascade}, into account. Refs. \cite{ZhuHammettPoP2010,ZhuDualCascade} work exactly with the original model and isolate the transfers in $\textbf{k}$ space and then are able to respect all of them individually. The canonical absolute equilibrium distribution corresponding to the present rugged invariants
is $\sim\exp\{-(\alpha_E \tilde{E} + \alpha_W \tilde{W})/2\}$
which gives the spectral density of $\tilde{E}$ and $\tilde{W}$:
$E(\textbf{k})\triangleq\langle \frac{\pi}{2} \beta(\textbf{k}) |\breve{\hat{g}}(\textbf{k},z_k)|^2 \rangle=\frac{\pi\beta(\textbf{k})}{2\pi\alpha_E\beta(\textbf{k})+4V^{-2}(\textbf{k}) \alpha_W J_1^{-2}[z_k]}$ and
$W(\textbf{k},z)\triangleq\langle 2V^{-2}(\textbf{k})J_1^{-2}(z)|\breve{\hat{g}}(\textbf{k},z)|^2 \rangle
=\frac{4}{\pi\alpha_E\beta(\textbf{k})J_1^2(z)\delta_{z,z_{(k)}}V^{2}(\textbf{k})+4\alpha_W}$.
Note that
\begin{eqnarray}\label{eq:we}
W(\textbf{k},z_k)=\frac{4}{\pi}V^{-2}(\textbf{k}) J_1^{-2}(z_k)E(\textbf{k})/\beta(\textbf{k}).
\end{eqnarray}
When $z\neq z_k$ (nondiagonal) or for the large $k$ limit, $W(\textbf{k},z)$ is equipartitioned as $1/\alpha_W$, and $E(\textbf{k})$ tends to condensate at the lowest modes of module $k_{min}$ for negative $\alpha_E$ with increasing $V^{-2}(\textbf{k}) J_1^{-2}(z_k)/\beta(\textbf{k})$. Such equilibria should be the states to which the system tend to relax and are relevant to the turbulence with collisions as simulated in LPT, according to Kraichnan \cite{KraichnanDualCascade}.

When $V^{-2}(\textbf{k})J^{-2}_1(z_k)/\beta(\textbf{k})$in Eq. (\ref{eq:we}) is monotonic (increasing for the present case) with $k$, then the Fj{\o}rtoft analysis for the constraints on the isolated transfers can be carried over \textit{mutatis mutandis}. LPT replaces $J_1^{-2}(z_k)$ with $z_k=V(\textbf{k})k$ which might not be that inaccurate in some limits or with some approximations, but, $V(\textbf{k})$ being not eligible to be normalized as done in LPT's Eq. (5) (see the third paragraph of this Comment), $V^{-2}(\textbf{k})J^{-2}_1(z_k)/\beta(\textbf{k})$ may not increase with $k$ if $V(\textbf{k})$ is not appropriately chosen: LPT's application of Fj{\o}rtoft argument is then flawed even for large $k$ where $\beta(\textbf{k})$ is approximated as a constant as they applied.

LPT neither respected the exact Eqs. (\ref{eq:gk}) and (\ref{eq:we}) nor treated the approximations appropriately, but the most important conceptual point is that the Fj{\o}rtoft constraint can not predict the turbulence transfer directions (as the sign of $\Delta E$ is undetermined so that the arrows in  Fig. 1 (a) of LPT could be reversed simultaneously, not to mention that the turbulence transfers are generally not isolated, but with dissipation and/or pumping, as the Fj{\o}rtoft analysis assumes) which however may be assisted by, among others, the tendency of relaxation to the absolute equilibria.

\smallskip

\noindent
Jian-Zhou Zhu\\
\indent {\small
Department of Modern Physics, \\\indent University of Science and Technology of China, \\\indent
 230026  Hefei, Anhui, China\\}

\end{document}